\documentclass[preprint,12pt]{elsarticle}

\usepackage{graphicx}
\usepackage{amssymb}
\PassOptionsToPackage{hyphens}{url}\usepackage{hyperref}

\usepackage{lineno}

\journal{}

\begin{document}

\begin{frontmatter}


\title{Non-constant hazard ratios in randomized controlled trials with composite endpoints}



\author{Jordi Cort{\'e}s\corref{cor1}\fnref{label1}}
\fntext[label2]{Department of Statistics and Operations Research\unskip, Universitat Polit{\`e}cnica de Catalunya\unskip, Jordi Girona, 31\unskip, Barcelona\unskip, 08034\unskip, Spain}
\cortext[cor1]{jordi.cortes-martinez@upc.edu}
%\ead{jordi.cortes-martinez@upc.edu}
%\ead[url]{https://jordi-cortes.netlify.com/}
\author{Mois{\'e}s G{\'o}mez Mateu\fnref{label1}}
%\ead{moises.gomez.mateu@upc.edu}
\author{KyungMann~Kim\fnref{label2}}
\fntext[label2]{School of Medicine \& Public Health\unskip, University of Wisconsin,\unskip Madison, \unskip Wisconsin,\unskip USA}
%\ead{kyungmann.kim@wisc.edu}
\author{Guadalupe G{\'o}mez Melis\fnref{label1}}
%\ead{lupe.gomez@upc.edu}

%\address{Barcelona, Spain}

%% Abstract -----------------------------------
\begin{abstract}
The hazard ratio is routinely used as a summary measure to assess the treatment effect in clinical trials with time-to-event endpoints. It is frequently assumed as constant over time although this assumption often does not hold. When the hazard ratio deviates considerably from being constant, the average of its plausible values is not a valid measure of the treatment effect, can be clinically misleading and common sample size formulas are not appropriate.

In this paper, we study the hazard ratio along time of a two-component composite endpoint under the assumption that the hazard ratio for each component is constant.

This work considers two measures for quantifying the non-proportionality of the hazard ratio: the difference $D$ between the maximum and minimum values of hazard ratio over time and the relative measure $R$ representing the ratio between the sample sizes for the minimum detectable and the average effects. We illustrate $D$ and $R$ by means of the ZODIAC trial where the primary endpoint was progression-free survival.

We have run a simulation study deriving scenarios for different values of the hazard ratios, different event rates and different degrees of association between the components. We illustrate  situations that yield non-constant hazard ratios for the composite endpoints and consider the likely impact on sample size.

Results show that the distance between the two component hazard ratios plays an important role, especially when they are close to 1.  Furthermore, even when the treatment effects for each component are similar,  if the two-component hazards are markedly different, hazard ratio of the composite is often non-constant.
\end{abstract}

%% Keywords -----------------------------------

\end{frontmatter}

%\linenumbers

\newpage

\section{Introduction}
The selection of the primary endpoint (PE) is a key decision which should be made at the first stage of a study. The combination of two or more outcomes in randomized controlled trials (RCTs) is often considered for the evaluation of the efficacy of new treatments and interventions  as it could  represent  a more comprehensive clinical picture \citep{STANLEY2007}. Composite endpoints (CEs), defined as the union of several outcomes, are extensively used when designing a clinical trial. In time-to-event studies, CE refers to the elapsed time from randomization until the earliest observation among its components. For example, it is common in oncological trials to use progression free survival (PFS): this outcome is defined as the time elapsed between randomization and objective tumor progression (OTP) or death from any cause, whichever occurs first \citep{Saad_2009}. Also, in cardiovascular trials, major adverse cardiac event (MACE) is generally defined as a composite endpoint that includes the time to cardiovascular death, myocardial infarction, stroke and target vessel revascularization \citep{GOMEZ_GOMEZMATEU_DAFNI_CIRC_14}.

When evaluating  the treatment effect on a time-to-event endpoint in an RCT, the hazard ratio (HR) is routinely used as a summary measure of treatment effect. When the PE is a CE, the hazard ratio of the CE,  $HR(t)$, might vary over time. In those cases where $HR(t)$ is approximately constant,  its average over time may appropriately capture the relative treatment effect between arms. However, otherwise, it  should not be understood as an average \citep{HERNAN_2010} and the common formulae to calculate sample sizes are not valid \citep{Schemper_2009}.

Nowadays more studies are encountered with non-constant hazard ratios and several reasons may explain why \citep{Halperin_1968}: i) many new therapies being evaluated have different modes of action, for example, the effect of the intervention might persist during the treatment period but diminishes afterwards, ii) phase III trials are much larger and therefore there are more chances to detect non-proportional hazards (NPH) \citep{Royston_2014}, and iii) with the intention of testing smaller treatment effects, composite endpoints are used more often \citep{KLEIST_2006} leading often to NPH. Indeed, it is proved that even under the assumption that for each component the  hazards are proportional, the resulting hazards for the CE are not necessarily proportional \citep{GOMEZ_2011_PROC}.

This paper discusses under which circumstances the hazard ratio of a CE may potentially result in greater departure from constancy and its impact on sample size. We base our evaluation on several simulated scenarios that represent realistic clinical trial situations. We illustrate the problem by means of a case study. Specifically, in this work, we provide: 1) two measures ($D$ and $R$)  as indicators of the non-proportionality of the hazards of the CE; 2) a list of factors (prioritizing according to their importance) that have an impact on the non-proportionality and consequently, on the sample size calculation; and 3) a threshold for the $R$ indicator above which the premise of the proportionality of the hazards is not reasonable. The final goal is to provide informed guidelines and warnings about the use of a constant HR when the deviation from constancy is remarkably high.

\section{Material and Methods}

\subsection{Notation and settings}
Consider an RCT comparing a new therapy  (group $g=1$) versus the standard therapy or control ($g=0$) with respect to a primary composite endpoint ${\cal E}_*$ where ${\cal E}_*$ is the union of  two different endpoints,  ${\cal E}_1$ and ${\cal E}_2$, for instance, overall survival (OS) and OTP.
We assume that the components forming the CE are relevant  for the objectives of the trial and their combination is clinically meaningful. Furthermore, the new therapy is supposed to be effective in the same direction for both endpoints, for instance, reducing the number of events.

In such an RCTs, individuals are followed until the event of interest  (${\cal E}_1$ or ${\cal E}_2$) or until the end of the study, whichever occurs first.  If we denote by $T^{(g)}_1$ and $T^{(g)}_2$ the times to ${\cal E}_1$ and ${\cal E}_2$, respectively, for individuals from group $g$, the time until the occurrence of ${\cal E}_* $, in group $g$, consisting of the earlier occurrence of ${\cal E}_1$ or ${\cal E}_2$, is denoted by $T^{(g)}_*$. Denote by  ${\rm HR_1}$ and ${\rm HR_2}$, the hazard ratios for ${\cal E}_1$ and ${\cal E}_2$, respectively, and assume that both are constant, as commonly done and denote by  ${\rm HR}_*(t)$, the hazard ratio for ${\cal E}_* $. We assume that ${\rm HR_k}<1$ ($k=1,2$)  implying that the new therapy reduces the risk of  both events  ${\cal E}_1$ and ${\cal E}_2$.

In order to study and characterize ${\rm HR}_*(t)$, the hazard ratio for ${\cal E}_* $, we need the following functions and parameters for $g=0, 1$: i) a joint distribution between $T^{(g)}_1$ and $T^{(g)}_2$, constructed using Frank's copula \citep{TrZi05}; ii) the marginal probability distributions for $T^{(g)}_1$ and $T^{(g)}_2$ chosen from a Weibull family because of its flexibility to represent different life-time data scenarios, allowing increasing, constant (exponential model) and decreasing hazard functions; iii) the probabilities $p_1^{(0)}$ and  $p_2^{(0)}$ of observing events ${\cal E}_1$ and ${\cal E}_2$ during follow-up up to $\tau$ in the control group $(g=0)$; iv)  the constant hazard ratios ${\rm HR_1}$ and ${\rm HR_2}$ for ${\cal E}_1$ and ${\cal E}_2 $, respectively,  and v)  the measure of association between $T_1^{(g)}$ and $T_2^{(g)}$, given by means of  Spearman's  rank correlation $\rho^{(g)}$ -we assume   $\rho^{(0)}=\rho^{(1)}=\rho$ . The reader is referred to \citep{GOMEZ_LAGAKOS_2013} for technical details concerning the computation of  ${\rm HR}_*(t)$.

In our setup we are considering that one of the events is fatal, say ${\cal E}_1$, hence  ${\cal E}_2$ would not be observed if it would  happen after  ${\cal E}_1$. We are therefore in a competing risks situation where the parameters governing the marginal distribution for  $T_2^{(g)}$ have to be  constrained to the  observed values of $T_2^{(g)}$ prior to $T_1^{(g)}$, that is, the probability of observing endpoint  $T_2^{(g)}$ is calculated as $p^{(g)}_2={\rm Prob}\{T^{(g)}_2<\tau, T^{(g)}_2<T^{(g)}_1\}$ where $\tau$ is the common censoring time corresponding to the end of the study.

\subsection{Non-Proportional Hazards Indicators}

If the effect of the treatment markedly changes during the period of follow-up, then the hazard ratio may vary over time. In order to capture the treatment effect in such cases, the options that the researchers may consider are: 1) the maximum value $MHR_*$ of the hazard ratio $HR_*(t)$, representing the minimum detectable effect; 2) the minimum value $mHR_*$ of the hazard ratio $HR_*(t)$, representing the maximum detectable effect; 3) the average value $aHR_*$ of the hazard ratio $HR_*(t)$ \citep{Kalbfleisch_1981}; or 4) other constant summaries \citep{Schemper_2009}. While $MHR_*$ and $mHR_*$ are never representative of the treatment effect when the hazard ratio varies over time, $aHR_*$ could be a valid summary of the treatment effect if the $HR_*(t)$ is reasonably constant.

Aiming to quantify the non proportionality of the $HR_*(t)$  we propose two measures: i)  the absolute difference between the maximum ($MHR_*$) and the minimum ($mHR_*$) values of the $HR_*(t)$ denoted by $D$:

\begin{equation}\label{rmeasure}
D=MHR_*-mHR_*
\end{equation}
and ii)  for a given significance level $\alpha$ and power,  the ratio between the sample size of the minimum detectable effect, $n_{MHR_*}$ and the  sample size of the average effect $n_{aHR_*}$, denoted by $R$:

\begin{equation}\label{Rmeasure}
R=\frac{n_{MHR_*}}{n_{aHR^*}}
\end{equation}

These two  measures complement each other: while $D$ provides a first intuitive quantification of how far the treatment effect is from being constant over time and of the impact of the non-proportionality on the effect size,  the impact of the non-proportionality on  the sample size is captured by $R$ with larger values of $R$ implying larger sample sizes. For the same distance $D$ between $MHR_*$ and $aHR_*$, the measure $R$ takes larger values when $aHR_*<MHR_*$ is closer to 1.

In the results section, their interpretation will be expanded.

\subsection{Sample size and R}

Following \citep{SCHOENFELD81,MACHIN1997} and assuming the same censoring rate for each group, at a one-tailed significance level $\alpha$ the number of events $(e_{h})$ and sample size $(n_{h})$ required to detect a treatment effect $\mbox{h}<1$ with power $1-\beta$ 
are given by:
\begin{equation}\label{Events}
e_{h}=4(z_{\alpha}+z_{\beta})^2/(\log \mbox{h})^2
\end{equation}
\begin{equation}\label{SS}
n_{h}=2 e_{h}/(  p^{(0)} + p^{(1)}  )
\end{equation}

\noindent where $\beta$ is the probability of type II error; $z_\alpha$ and $z_\beta$ are the standard normal quantiles corresponding to $\alpha$ and $\beta$, respectively; and $p^{(g)}$ is the probability of observing the event in group $g$ during the study.  It can be shown that
the relative measure $R$ is

\begin{equation}\label{Rmeasure}
R=\left( \frac{\log(aHR_*)}{\log(MHR_*)} \right)^2=\frac{n_{MHR_*}}{n_{aHR^*}}
\end{equation}

\noindent
$n_{aHR^*}$ provides the sample size based on the average value of $HR_{*}(t)$ as treatment effect while $n_{MHR_*}$ is an upper bound of the sample size obtained for the minimum detectable effect. If $HR_{*}(t)$ is constant, $n_{aHR^*}$ should be close to $n_{MHR_*}$ and hence $R$ close to 1. Otherwise, $R$ values away from 1 indicate that conventional sample size formulas for constant $HR(t)$ can not be used.\\

\subsection{Simulation Study Settings}

The behaviour of the hazard ratio $HR_*(t)$ for the composite of two endpoints is studied for a variety of scenarios based on different parameter values as shown in Table \ref{Table_scenarios}. We have chosen parameter values for simulation that represent realistic scenarios when designing an RCT \citep{GOMEZ_LAGAKOS_2013}.

\begin{table}[h!]
	\centering
	\begin{tabular}{c|cccc}
			\textbf{Parameters} &&&&\\
			\hline\hline
			$p_1^{(0)},p_2^{(0)}$ &0.1&0.3&0.5&\\
			${HR}_{1}, {HR}_{2}$  &0.6&0.7&0.8&0.9\\
			$\rho$                &0.1&0.3&0.5&\\
			
			\hline
			Distribution            &(Decr.\ hazards) &(Exponential) &(Incr.\ hazards)&\\
			$\beta_{1} , \beta_{2} $&0.5              &             1&               2&\\
			
			\hline\hline
			Number                     &&&&\\
			of scenarios&\textbf{3 888}&&&\\
			\hline
	\end{tabular}
		\caption{Parameter setting to generate the simulation scenarios. $p_k^{(0)}$ are the  probabilities of observing each component ${\cal E}_{k}$ in the control group; $HR_{k}$ are the constant cause-specific hazard ratios for each event ${\cal E}_k$; $\rho$ stands for the Spearman's rank correlation; $\beta_k$ are the shape parameters of the Weibull distribution for ${\cal E}_k$. The total number of scenarios are derived from all possible combinations $(3^2 \times 4^2 \times 3 \times 3^2)$.}
    \label{Table_scenarios}
\end{table}

The probabilities $p_1^{(0)}$ and $p_2^{(0)}$ of observing each component event in the control group have been taken between 0.1 and 0.5; the cause-specific hazard ratios $HR_1$ and $HR_2$ of each component represent large to small treatment effects (from 0.6 to 0.9); the correlations between the time to each endpoint have been selected from low to moderate (from $\rho=0.1$ to $\rho=0.5$); and the times until the component endpoints (${\cal E}_k, \ k=1,2)$ have been modeled according to Weibull distributions with constant $(\beta_k=1)$, decreasing $(\beta_k=0.5)$ or increasing hazards $(\beta_k=2)$. For simplicity, the follow-up time has been set constant and equal to 1 in all scenarios without involving any loss of generalisability in the results.

\section{Results}

\subsection{ZODIAC trial}

ZODIAC trial \citep{Herbst_2010} compared the efficacy of the Vandetanib plus docetaxel versus docetaxel as second-line treatment in patients with advanced non-small-cell lung cancer. Statistically significant differences were found in the composite primary endpoint, PFS, taken as the union of OS and OTP.

\begin{figure}[h!]
	\centering
	\resizebox*{12.5cm}{!}{\includegraphics[width=1\linewidth]{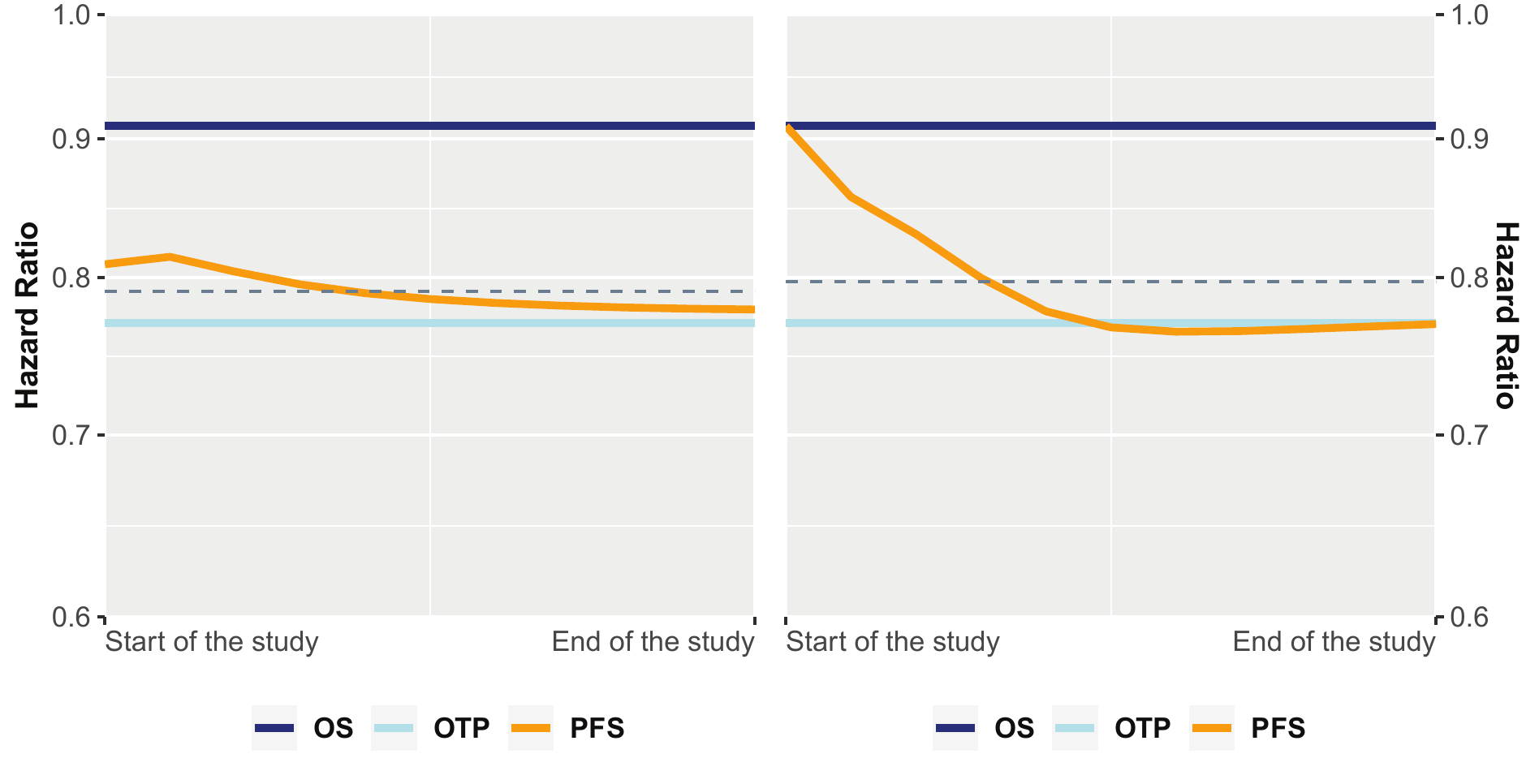}}
	\caption{HRs over time in ZODIAC trial under two different scenarios. Dark and light blue lines are the cause-specific hazard ratios for Overall Survival (OS) ($HR_{OS}=0.91$) and Objective Tumor Progression (OTP) ($HR_{OTP}=0.77$), respectively. Orange line represents the hazard ratio for the composite endpoint ($HR_*(t)$) over time. In both plots, the probabilities of OS and OTP in the control group are $p_{OS}^{(0)}=0.59$ and $p_{OTP}^{(0)}=0.74$, respectively, and a moderate correlation ($\rho=0.5$) is assumed between the  events. The difference between the two plots is due to the different distributions of the times $T_{OS}$ and $T_{OTP}$: while in the left plot, an exponential distribution is assumed for both  events, the right plot assumes a Weibull distribution with increasing hazards  for OTP  and an exponential distribution for OS.}
	\label{Plots_HR}
\end{figure}

The reported cause-specific  HRs for each component were 0.91 (OS) and 0.77 (OTP) and the probabilities of observing each component event in the control group were 0.59 and 0.74, respectively. Furthermore, we know that the estimated HR of the composite event, PFS, was 0.79. From these data, we  explore a couple of situations with different resulting interpretations. Figure \ref{Plots_HR} shows the shape of the hazard ratio for  PFS,  $HR_*(t)$, during the study in two different scenarios. Assuming a moderate association ($\rho=0.5$) between OS and OTP, if the marginal times to death and OTP follow exponential distributions (constant hazards), then  $HR_*(t)$  would be almost constant fluctuating in a narrow range from 0.78 to 0.81 over time (orange curve in the left plot) and with  $aHR_*$, average of  $HR_*(t)$, during follow-up, equal to 0.79  which is consistent with the reported  hazard ratio.  $aHR_*=0.79$ represents a $21\%$ relative reduction in the risk of death or OTP in the treatment arm (Vandetanib plus docetaxel) as compared with the control arm (docetaxel). In this situation, the  average  $aHR_*$ would be a good approximation of the treatment effect at any instant of time. However, if as it often does, the hazards of OTP are increasing, regardless of the treatment arm, then the $HR_*(t)$ for PFS could vary markedly (along time) taking extreme values from 0.76 up to 0.91 (orange curve in the right plot). Obviously, in this case, although the  value of $aHR_*$ is the same as  in the previous scenario (0.79), it is not representative at all of the treatment effect over time. 

The assumed distributions for  the individual components significantly affect the sample size calculations in both situations of Figure \ref{Plots_HR}. If both distributions are exponential, implying constant hazards (Figure \ref{Plots_HR}, left), despite the small difference ($D=0.03$) between the $MHR_*$ (0.81) and $mHR_*$ (0.78), the implication on the sample size is significant: the indicator $R$ is equal to 1.25, implying an increase of 25\% in the number of patients. However, it is even worse if the hazards increase over time in the OTP endpoint (Figure \ref{Plots_HR}, right): now, the difference in the HRs ($D=0.15$) implies serious concerns in the sample size since $R$ increases up to 6.25 leading to an increase of 525\% if minimum detectable effect is considered for the sample size calculations.

In the ZODIAC trial, 1,176 events were considered enough to reach a power of 0.9 to detect a HR equal to or less than 0.8 with a 0.05 significance level. The study obtained a p value $=0.0001$ for PFS with a larger effect size than that specified in the sample size calculations, which facilitated reaching statistical significance. Panel A in Figure 2 of \citep{Herbst_2010}  shows the Kaplan-Meier curves for PFS; despite their similar behavior during the first two months for both arms, the subsequent gradual separation does not point out a violation of the proportional hazards assumption. 

All these results have been obtained using the CompARE platform  (\url{http://cinna.upc.edu:3838/compare/CompARETimeToEvent/}), which is a website specifically created for the design of clinical trials with composite endpoints. It allows, among others, to perform sample size calculations and to draw the $HR_*(t)$ depending on some input parameters, such as the cause-specific hazard ratios, the probabilities of observing each event or the correlation between outcomes.

\subsection{Understanding D and R measures}

\begin{figure}[!ht]
	\centering
	\resizebox*{12.5cm}{!}{\includegraphics[width=1\linewidth]{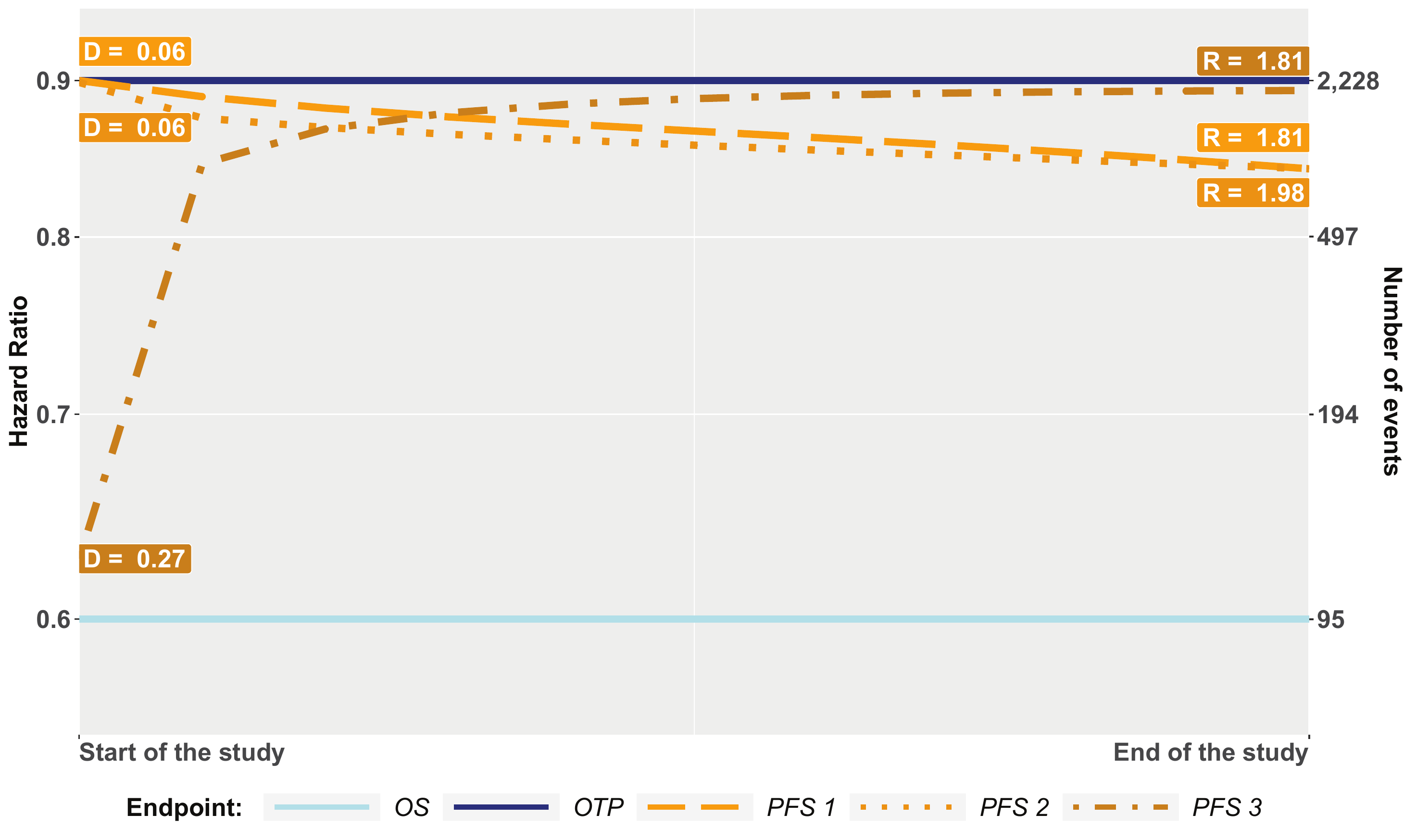}}
	\caption{Behaviour of the $HR_*(t)$ (orange) for PFS in three different scenarios. $HR_{OS}$ and $HR_{OTP}$ are the cause-specific hazard ratios of each component: ${\cal E}_1= {\rm death}$ (dark blue) and ${\cal E}_2={\rm OTP}$ (light blue). The probabilities of observing the events during the study in the control group are $p_{OS}^{(0)}=0.15$ and $p_{OTP}^{(0)}=0.5$, respectively, with a correlation coefficient $\rho=0.3$. The three scenarios are characterized by the hazards behaviour over time in the distributions of $T_{OS}$ and $T_{OTP}$: a) increasing (OS)/constant (OTP) -- dotted line--; b) constant (OS)/decreasing (OTP) -- dashed line--; and c) decreasing (OS)/constant (OTP) -- semidotted line--. Left vertical axis represents the hazard ratio and right vertical axis represents the needed sample size at a one-sided significance level $\alpha=0.05$ with power $1-\beta=0.8$ depending on the corresponding HR at left. Indicators of non-proportionality ($D$ and $R$) are placed next to the line representing each situation.}
	\label{r_R_2}
\end{figure}

Figure \ref{r_R_2} shows the $HR_*(t)$ function for three different settings (orange lines). These scenarios only differ in the behaviour of the hazard of the time to each component (see legend of Figure \ref{r_R_2}). We observe that identical departures indicated by $D$ can have different implications. For instance, the dashed and dotted lines have similar behaviour over time and the same extreme values ($mHR_*=0.84$ and $MHR_*=0.90$) and thus, the same absolute difference $D=0.06$; however, the measure $R$ takes distinct values in both scenarios ($R=1.81$ and $R=1.98$), indicating that the required number of patients using  $MHR_*$ is respectively $81\%$ and $98\%$ larger than if $HR_*(t)$ is summarized by the average $aHR_*$. 

The right axis of Figure \ref{r_R_2} represents the number of events to achieve a power=0.8 using the formula (\ref{Events}) at a one-sided significance level $\alpha=0.05$ using the corresponding HR indicated on the left axis. It can be seen that the number of events substantially increase in the range from a  hazard ratio $h=0.8$ to $h=0.9$. For this reason, minor fluctuations in this area have great repercussions on the sample size. For instance, in both, the dashed and dotted lines ($PFS 1$ and $PFS 2$), the required number of events based on the $MHR_*$ is around 2,200 while based on the average hazard ratio $aHR_*$, only about 1,200 and 1,100 events are required, respectively. 

Following the example in Figure \ref{r_R_2}, we depict the $HR$ for PFS with the semi-dotted line ($PFS 3$) in this case, with a decreasing hazard function for OS, and a constant hazard for OTP. The behaviour of $HR_*(t)$ is markedly different from earlier to later times of follow-up, producing a much larger absolute difference $D=0.27$. If we now compare the dashed and semi-dotted lines, we  observe that the same $R=1.81$ leads to a noticeable difference between ranges ($D=0.06$ and $0.27$, respectively). As it will be seen later, the  choice of the distribution of each component in the sample size determination is paramount: small differences of the anticipated value for $HR$ could dramatically change the required number of events (specially for hazard ratios close to 1, as we will discuss later) and vice versa.

In the simulation study presented next, we will study under which situations there is a greater departure of constancy for the $HR_*(t)$  and the  consequences in the calculation of the sample size.

\subsection{Simulation Study}

We will assess under which scenarios the premise of proportional hazards will be more compromised in terms of sample size implications, and for this reason the $R$ indicator will be used during this section. Under the parameter setting in Table \ref{Table_scenarios}, Table \ref{Table_rR_betas_HRs} reports on the influence of the input parameters (treatment effects, hazard behaviour and correlation) on the measure $R$.

\begin{table}[h!]
	\centering
	\begin{tabular}{l|ccc} 
			&&\textbf{R}&\\
			&\textbf{Minimum}&\textbf{Median}&\textbf{Maximum}\\
			\hline\hline
			T\textbf{reatment effect}\\
			$HR_1 = HR_2$		    &1&1.05&1.35\\
			$|HR_1 - HR_2| = 0.1$	&1&1.20&3.49\\
			$|HR_1 - HR_2| = 0.2$	&1&1.49&8.18\\				
			$|HR_1 - HR_2| = 0.3$	&1&2.06&15.65\\
			\hline \hline
			\textbf{Laws of each component}\\			
			Both decreasing hazards ($\beta_1=\beta_2=0.5$)		    &1	 &1.04&1.23\\
			Both constant  ($\beta_1=\beta_2=1$)			        &1	 &1.04&1.28\\	
			Both increasing hazards ($\beta_1=\beta_2=2$)		    &1	 &1.06&1.44\\
			Different behaviour in hazards ($\beta_1 \neq \beta_2$)	&1.01&1.39&15.65\\	
			\hline \hline
			\textbf{Correlation}\\
			Weak ($\rho=0.1$) 		&1		&1.07&14.97\\
			Mild ($\rho=0.3$)		&1.01	&1.13&15.19\\
			Moderate ($\rho=0.5$)	&1.01	&1.18&15.65\\
			\hline \hline
			\textbf{Global}		    &\textbf{1}&\textbf{1.15}&\textbf{15.65}\\
			\hline			
	\end{tabular}
	\caption{Range and median values of the $R$ indicator depending on several factors: i) absolute difference between the HRs of the components; ii) the laws of the times to each event; and iii) the correlation between these times.}
	\label{Table_rR_betas_HRs}	
\end{table}

We observe that when the treatment effects for each component are equal, i.e. $HR_1=HR_2$, the median value of $R$ is 1.05, meaning a $5\%$ of relative difference between the sample sizes required by $MHR_*$ and $aHR_*$, but the median increases up to 2.06 when the treatment effects markedly differ ($|HR_1-HR_2| = 0.3$). Therefore, greater concerns regarding the sample size arise when the treatment  effect in each component is noticeably different.

Furthermore, when the behaviour of the hazards is different in each endpoint ($\beta_1\ne\beta_2$), smaller effects, that is $HRs$ close to 1, imply higher values of the $R$ indicator (not shown in Table \ref{Table_rR_betas_HRs}). For instance, for an absolute difference of the $HRs$ equal to 0.1, the median of $R$ is 1.23 if the $HRs$ of each endpoint are 0.6 and 0.7, while it is 1.60 if these $HRs$ are 0.8 and 0.9.

When the probability distributions governing the time to each endpoint have the same hazard,
i.e., $\beta_1=\beta_2$, the median of $R$ under different settings is no more than 1.06. That is, if the event times have the same shape parameter, the  $HR_*(t)$ is relatively constant. It is not the case if two endpoints with different hazard functions are combined: the median departure from constancy is 1.39 and can be as large as 15.65. Our results do not point out a strong relationship between the $R$ values and the degree of association between component endpoints.

Figure \ref{R_scenarios} shows the complete distribution of the simulated $R$ values under various scenarios. Again, it can be observed that the relative measure $R$ depends strongly on the behavior of the  hazard function for each component. If the evolution over time of  the two endpoint hazards occurs in different directions  (i.e., one with increasing hazard and the other with decreasing one), then in almost all situations, the implications of considering a constant $HR(t)$ on the sample size can be much larger. Only if the treatment effects are similar in both outcomes, this fact would be mitigated.

In the remaining scenarios, when the probability distributions governing the time to each endpoint have the same hazard function, i.e., $\beta_1=\beta_2$, high $R$ values can be only obtained if treatment effects are very different ($|HR_1-HR_2| \ge 0.3$) and the correlation between components are moderate or high ($\rho \ge 0.5$). In all scenarios, high correlations and different treatment effects imply larger values of $R$.

\subsection{Criterion to assess non-proportionality}

\begin{figure}[!h]
	\centering
	\resizebox*{12.5cm}{!}{\includegraphics[width=1\linewidth]{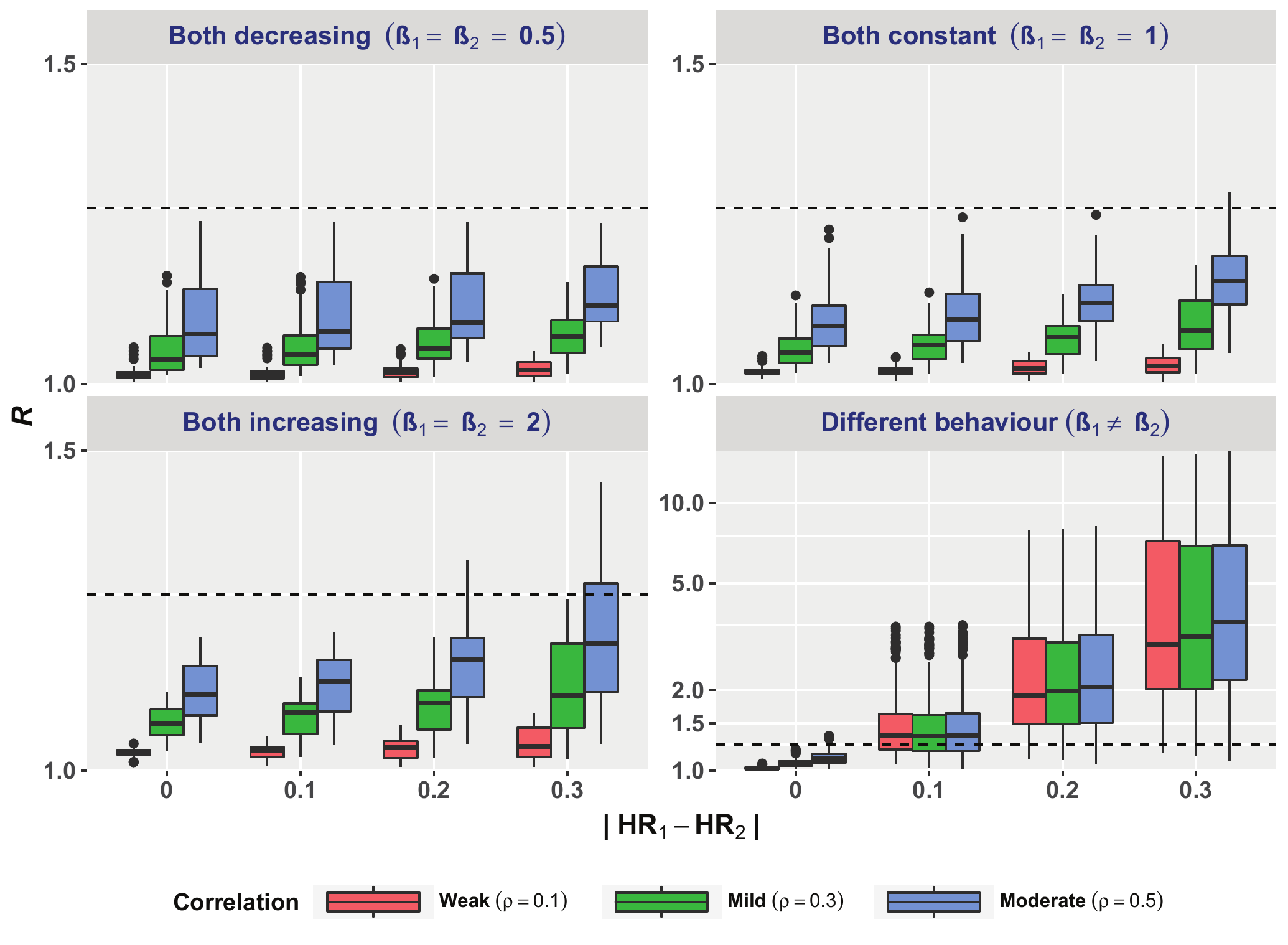}}
	\caption{Quantification of the impact of various factors on the sample size. $R$ indicator as function of the hazard of each endpoint (panels), the absolute difference of the HRs (horizontal axis) and the correlation between endpoints (color). Vertical axes represent the $R$ value in log-scale - note that the scale of the last panel is different from the others. The dashed horizontal lines at $R = 1.25$ in each panel represent the value from which the non-proportionality of the hazards is critical and should be taken into account in the sample size calculations.}
	\label{R_scenarios}
\end{figure}

In order to establish a criterion to decide when the average $aHR_*$ of $HR_*(t)$ is not a meaningful summary for the treatment effect on the composite endpoint, we have taken into account the following: we consider that the HR of the CE would be remarkably non-constant if the required sample size using  $MHR_*$ is $25\%$ larger than that using $aHR_*$. Based on this criterion, we recommend a threshold of $1.25$ (horizontal dashed line in each panel of Figure \ref{R_scenarios}) for $R$. It is worth mentioning that when the treatment effects and/or the distributions in each component are very different, it is very likely to exceed this threshold.

\section{Conclusions}
In randomized controlled trials it is common to use a hazard ratio as a summary to assess the treatment effect on the time to event outcomes and to base the sample size calculation on the constant hazard ratio defined in the alternative hypothesis. Since composite endpoints are frequently used as primary endpoints in many fields, we are concerned about whether it is reasonable  to assume a constant hazard ratio for the composite endpoint and use the standard formulae for sample size.

We have described  situations where the hazard ratio for a composite endpoint is far from being constant,  taking into account the behaviour of the  hazard function and the probability of observing the event of each component for the control group as well as the association between the component endpoints. We have seen that component endpoints arising from  different probabilistic distributions (e.g., one exponential and the other Weibull with increasing hazards) or very different treatment effect lead almost always to non-constant hazard ratios. In these scenarios, the average hazard ratio does not provide a meaningful measure of the treatment effect and alternatives such as the difference in percentiles at the end of the study or the restricted mean survival time, defined as the expected event-free time during the follow-up \citep{UNO_2015}, should be used instead.

Next, we discuss some specific issues of our work. First, we have considered scenarios with one terminal event, such as death. This is a critical consideration because of concerns related with competing risks. Despite the estimation of the marginal distributions of the competing outcomes is generally infeasible \citep{Tsiatis_1975}, we can define the marginal distribution for each component of the composite endpoint under the latent failure time model, that is, assuming potential marginal distributions for $T_1^{(g)}$ and $T_2^{(g)}$ and  obtain the joint distribution of  the two endpoints  by means of a  copula, even though we are aware that this is not the unique alternative. Nevertheless, we are not encountering any problem because $T^{(g)}_*$ ($g=0,1$) includes the death component, it is observable based on real data and the parameters governing the marginal distribution for $T_2^{(g)}$ are constrained to the  observed values of $T_2^{(g)}$ prior to  $T_1^{(g)}$. Second, the choice of Spearman's correlation as a measure of  association has been made for interpretative reasons. Other popular association measures, such as Kendall's $\tau$, could be used instead. Since there is a one-to-one correspondence between Kendall's $\tau$ and Spearman's $\rho$, the choice of one  versus the other  does not have any repercussion when using copulas to generate the joint distribution. Third, in all scenarios, we have defined a follow-up time equal to 1 just for convenience, but the findings can be extrapolated for any follow-up time provided that the parameters of the distributions are scaled accordingly. Fourth, we have set a limit value of $R = 1.25$ keeping in mind that the calculation of the sample size based on a constant $HR(t)$ with values above this threshold can be problematic. Obviously, this threshold should be decided by each researcher paying attention to the conditions described in this work that are more unfavorable to have a constant hazard ratio. Fifth and last, in the case of the ZODIAC Trial, the censoring comes from, on the one hand,  the end of the study and, on the other hand, the losses of follow-up. When performing the calculations of Figure \ref{Plots_HR}, only the end of study censoring have been considered.

Our work attempts to highlight the inappropriateness of the proportional hazards assumption for composite endpoints in several situations, especially when sample size is based on this erroneous premise. In those cases where this proportionality is not fulfilled, other alternative measures should be used and the design and statistical analysis of the clinical trial should change accordingly. This paper is not addressing aspects of the statistical analysis when the proportional hazards assumption does not hold.

We advocate for a thorough study of the possible patterns of the hazard ratio for the composite endpoints before deciding on the primary endpoint for efficacy and prior to any sample size estimation. The researcher must consider 1) if the behavior of the hazards are different among the components; 2) if the treatment effects are different in each component; and 3) if the occurrences of the events are somehow correlated. The more affirmative answers to these three questions, the more likely a non-constant treatment effect over time for the CE. For a quantitative study of the influence of these factors, one may use the CompARE platform (\url{http://cinna.upc.edu:3838/compare/CompARETimeToEvent/}). All these considerations should be included and discussed in the protocol of the study allowing trialists to evaluate whether the hazard ratio is an appropriate measure or whether other measures should be considered for the evaluation of the treatment effect.

\section*{Acknowledgements}

This work was partially supported by the Ministerio de Econom\'{i}a y Competitividad (Spain) [MTM2015-64465-C2-1-R (MINECO/FEDER)]; the Departament d'Economia i Coneixement de la Generalitat de Catalunya (Spain)[2017 SGR 622 (GRBIO)].

\newpage

%\section{References}

\newpage

\end{document}